\documentclass[twocolumn,english,showpacs,floatfix]{revtex4}

\usepackage[T1]{fontenc}
\usepackage[latin1]{inputenc}
\usepackage{amssymb}

\makeatletter

\usepackage{graphicx}
\usepackage{amssymb}
\usepackage{bm}

\usepackage{babel}
\makeatother
\begin{document}

\title{A self-consistent theory of atomic Fermi gases with a Feshbach resonance at
the superfluid transition}

\author{Xia-Ji Liu and Hui Hu}

\affiliation{\ ARC Centre of Excellence for Quantum-Atom Optics, Department
of Physics, University of Queensland, Brisbane, Queensland 4072, Australia}

\date{\today{}}

\begin{abstract}
A self-consistent theory is derived to describe the BCS-BEC crossover for a
strongly interacting Fermi gas with a Feshbach resonance. In the theory the
fluctuation of the dressed molecules, consisting of both preformed
Cooper-pairs and ``bare'' Feshbach molecules, has been included within a
self-consistent $T$-matrix approximation, beyond the Nozi\`{e}res and
Schmitt-Rink strategy considered by Ohashi and Griffin. The resulting
self-consistent equations are solved numerically to investigate the normal
state properties of the crossover at various resonance widths. It is found
that the superfluid transition temperature $T_c$ increases monotonically at
all widths as the effective interaction between atoms becomes more
attractive. Furthermore, a residue factor $Z_m$ of the molecule's Green
function and a complex effective mass have been determined, to characterize
the fraction and lifetime of Feshbach molecules at $T_c$. Our many-body
calculations of $Z_m$ agree qualitatively well with the recent measurments on the
gas of $^6$Li atoms near the broad resonance at $834$ Gauss. The crossover
from narrow to broad resonances has also been studied.
\end{abstract}

\pacs{03.75.Ss, 05.30.Fk, 74.20.-z}

\maketitle
\textit{Introduction}. --- The recently demonstrated strongly interacting
Fermi gases of $^{40}$K and $^6$Li near a Feshbach resonance have attracted
a lot of attention \cite{jin,ketterle,thomas04,grimm04,ens,chin,thomas05,vortex}. 
In these systems, because of the coupling between a pair of atoms in an open 
channel and a bound molecular state of a closed channel, it is possible to 
obtain an arbitrarily strong attractions between atoms by tuning the energy 
of the molecular state across the resonance with a magnetic field. It 
therefore provides us a unique opportunity to experimentally examine the 
crossover from a Bardeen-Cooper-Schrieffer (BCS) superfluidity to a 
Bose-Einstein condensate (BEC).

The use of the Feshbach resonance, however, significantly complicates the
theoretical analysis of current experimental results, due to the presence of
the bound molecular state. It has been argued that a two-channel (resonance)
model that describes explicitly both the open channel and closed channel
physics should be adopted as a minimum ingredient for the theory \cite
{karen,holland,timmermans}. In the model, the molecular bound state in the 
closed channel is treated as a featureless bosonic particle, and two essential
parameters, the threshold energy of molecules and the atom-molecule coupling
constant, are required to characterize the total system. This is in contrast 
to the traditional single-channel model, in which the closed channel or 
the ``bare'' Feshbach molecules are neglected. Thereby the model is 
characterized solely by a single parameter of $s$-wave scattering length $a$, 
\textit{i.e.}, the interatomic interaction is described by a contact potential with 
strength $4\pi \hbar ^2a/m$. The applicability of these models to current experiments, 
together with their equivalence in certain limits, are actively debated at present. 
A closely related subtle problem is the nature of Fermi superfluidity near the 
resonance, \textit{i.e.}, whether the superfluid transition is associated 
with small pairs that are molecular in character, or large objects akin to 
Cooper pairs.

In this paper we present a many-body theory of the two-channel model for an
interacting ultra-cold Fermi gas with a Feshbach resonance at and above the
superfluid transition temperature $T_c$. In the theory the strong
fluctuations of the preformed Cooper-pairs and of the Feshbach molecules
have been incorporated within a self-consistent $T$-matrix approximation,
similar to that considered by Haussmann for single-channel model \cite
{haussmann}. We then apply the theory to explore the normal phase of the gas
at BCS-BEC crossover, emphasizing the importance of the resonance width. The
purpose of the exploration is three-fold. \textbf{(1)} First, a reliable $T_c
$ is predicted. Till now there are strong evidences for the emergence of
superfluidity at the crossover \cite{jin,ketterle,thomas04,grimm04,chin,thomas05,vortex}. 
Precise experimental characterizations of the thermodynamic and dynamical properties 
of strongly interacting Fermi gases are also undergoing. In this respect, a reliable
theoretical prediction of $T_c$ of the crossover gas is, therefore, useful
for guiding future experimental investigations. Previous calculations of $T_c
$ is based on the so-called Nozières and Schmitt-Rink (NSR) scheme \cite
{nsr,griffin}, which is an analysis equivalent to the ladder approximation
using free fermionic Green's functions. In both weak and strong coupling
limits this approximation leads to correct results \cite{nsr}. However, at
the crossover it becomes less accurate as the fermionic quasiparticles are
far from being free particles. Our theory has the advantage of taking a
self-consistently dressed Green function in the $T$-matrix, and is expected
to give a reasonable $T_c$ than the NSR scheme. \textbf{(2)} Second, to
characterize the nature of pairs at the crossover, the fraction and lifetime
of the ``bare'' Feshbach molecules have been studied, by extracting the
residue factor $Z_m$ and the effective mass from the Green function of
molecules. The value of $Z_m$ can be directly determined in experiments
using the Stern-Gerlach selection technique to measure the magnetic moment
of the pairs \cite{grimm03}, or using the optical spectroscopy to project
out the singlet component of the pairs \cite{hulet}. Our calculations of $Z_m
$ emphasize the many-body effects at the crossover, in particular, on the
BCS side of the resonance. For an ultracold gas of $^6$Li atoms at the broad
resonance $B_0=834$ \textrm{G}, we find a reasonable agreement between our
theoretical predictions and the recent spectroscopy measurements by Hulet 
\textit{et al}. \cite{hulet}. \textbf{(3)} Finally, the crossover from a
narrow to a broad resonance with increasing atom-molecule coupling strengths
has been investigated.

\textit{Self-consistent $T$-matrix approximation}. --- We consider a
homogeneous gas of fermionic atoms in two different hyperfine states in the
vicinity of a Feshbach resonance. In the grand canonical ensemble the gas of 
$N$ atoms is described by an atom-molecule Hamiltonian \cite{karen,holland,timmermans},
\begin{eqnarray}
\mathcal{H} &=&\sum_{\mathbf{k}\sigma }\left( \epsilon _{\mathbf{k}}-\mu
\right) c_{\mathbf{k}\sigma }^{+}c_{\mathbf{k}\sigma }+U_{bare}\sum_{\mathbf{%
kk}^{\prime }\mathbf{q}}c_{\mathbf{k+q}\uparrow }^{+}c_{\mathbf{k}^{\prime }-%
\mathbf{q}\downarrow }^{+}c_{\mathbf{k}^{\prime }\downarrow }c_{\mathbf{k}%
\uparrow }  \nonumber \\
&&+\sum_{\mathbf{q}}\left( \epsilon _{\mathbf{q}}/2+\nu _{bare}-2\mu \right)
b_{\mathbf{q}}^{+}b_{\mathbf{q}}  \nonumber \\
&&+g_{bare}\sum_{\mathbf{kq}}\left( b_{\mathbf{q}}^{+}c_{\mathbf{k+q/2}%
\uparrow }c_{-\mathbf{k+q/2}\downarrow }+h.c.\right) .  \label{hami}
\end{eqnarray}
Here the two hyperfine states are denoted as pseudospins $\sigma =\uparrow
,\downarrow $, with $N_{\uparrow }=N_{\downarrow }=N/2$. $c_{\mathbf{k}%
\sigma }^{+}$ and $b_{\mathbf{q}}^{+}$ represent the creation operators of
an atom with the kinetic energy $\epsilon _{\mathbf{k}}=\hbar ^2\mathbf{k}%
^2/2m$ and of a Feshbach molecule with the dispersion $\epsilon _{\mathbf{q}%
}/2+\nu _{bare}=\hbar ^2\mathbf{q}^2/4m+\nu _{bare}$, respectively. The
magnetic detuning or the threshold energy $\nu _{bare}$ is a key parameter
in this model, and can be conveniently adjusted by an external magnetic
field $B$. It changes sign across the resonance from above and enables the
formation of stable molecules. As the molecule is made of two atoms, a
chemical potential $\mu $ is introduced to impose the conservation of the
total number of the bare Fermi atoms, \textit{i.e.}, $\left\langle \mathcal{N%
}\right\rangle =\left\langle \sum_{\mathbf{k}\sigma }c_{\mathbf{k}\sigma
}^{+}c_{\mathbf{k}\sigma }\right\rangle +\left\langle \sum_{\mathbf{q}}b_{%
\mathbf{q}}^{+}b_{\mathbf{q}}\right\rangle =N$. Moreover, two contact
interactions have been adopted for the nonresonant (or back-ground)
atom-atom scatterings and the atom-molecule coupling, as parameterized by $%
U_{bare}$ and $g_{bare}$, respectively.

As usual, the use of contact interactions leads to an (unphysical)
ultraviolet divergence, and requires a renormalization that expresses the
bare parameters $U_{bare}$, $g_{bare}$, and $\nu _{bare}$, in terms of the
observed or renormalized values $U_0=4\pi \hbar ^2a_{bg}/m$, $g_0$, and $\nu
_0=\Delta \mu \left( B-B_0\right) $, where $a_{bg}$ is the back-ground
s-wave scattering length for the atoms, $g_0$ is associated with the width
of resonance, and $\Delta \mu $ is the magnetic moment difference between
the atomic and the bound molecular states. The renormalization can be done
by solving either the Lippmann-Schwinger scattering equations or the vertex
functions in the two-body limit. The outcome is very transparent in physics:
one can just replace in everywhere the bare parameters by their
corresponding observable values. In addition, the divergent Cooper pair or
two-particle propagator (defined later) could be regularized as $\Pi (\mathbf{q},i\nu
_n)\rightarrow \Pi (\mathbf{q},i\nu _n)-\sum_{\mathbf{k}}1/\left( 2\epsilon
_{\mathbf{k}}\right) $. In Ref. \cite{drummond}, it is stated that this
procedure yields the correct expression for the molecular binding energy,
and therefore incorporates the exact two-body scattering process. A similar
conclusion has also been arrived by Falco and Stoof \cite{stoof}.

To relate the value of $g_0$ to experimentally known parameters, we compare
the expression for the effective interactions between atoms in the
low-density limit, $U_{eff}=U_0+g_0^2/\left( -\nu _0\right) $, with the
s-wave scattering length of the atoms as a function of magnetic field, $%
a(B)=a_{bg}\left[ 1+\Delta B/(B-B_0)\right] $, where $\Delta B$ is the width
of resonance. Using the relation $U_{eff}=4\pi \hbar ^2a(B)/m$ and
eliminating $U_0$ and $\nu _0$, we have that $g_0=2\hbar \left[ \pi \left(
-a_{bg}\Delta B\Delta \mu \right) /m\right] ^{1/2}$. For the $^6$Li atoms at
the broad resonance $B_0=834$ \textrm{G }of interest here, $\Delta B\approx
300$ \textrm{G}, $\Delta \mu \approx 2\mu _B$, and $a_{bg}\approx -1405a_0$ 
\cite{grimm05}, where $\mu _B$ is the Bohr magneton and $a_0=0.529$ \textrm{Å} 
is the Bohr radius.

\begin{figure}
\includegraphics[%
  width=5cm,angle=-90]{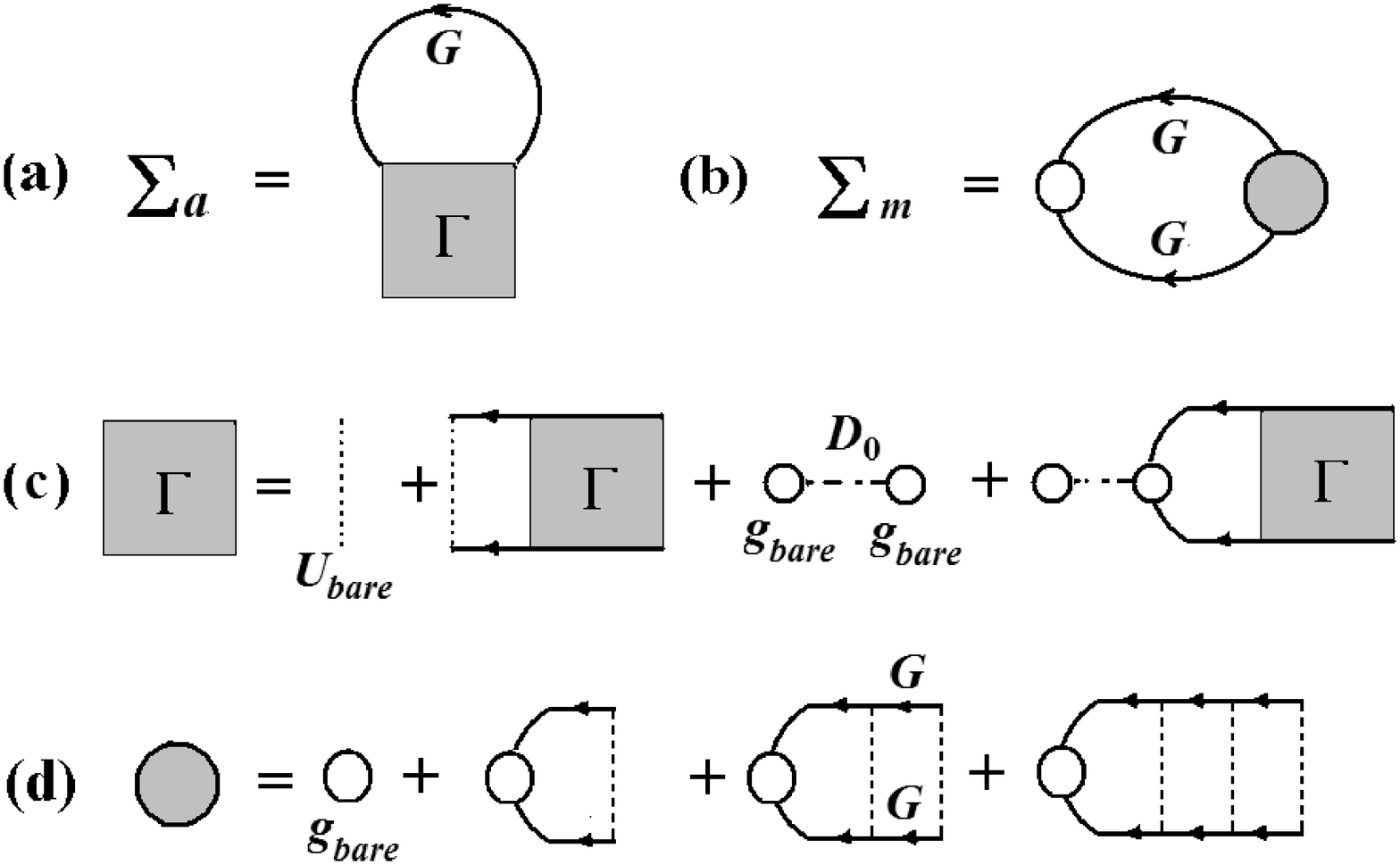}

\caption{Diagrammatic representations of the atomic self-energy (a), the
molecular self-energy (b), the Bethe-Salpeter equation for the four-point
vertex function (c), and the atom-molecule vertex function (d). The solid
(dot-dashed) line represents $G$ ($D_0$), the dotted line describe $U_{bare}$,
and the empty circle is the bare atom-molecule coupling $g_{bare}$.}

\label{fig1}
\end{figure}

We solve the atom-molecule Hamiltonian (\ref{hami}) in the self-consistent $T$
-matrix approximation by summing up infinite ladder diagrams involving $U_{bare}$ 
and particle-particle bubble diagrams involving $g_{bare}$. We
consider here only the normal phase of the atomic gas. The generalization to
the superfluid phase is more interesting (and, of course, more complicated),
and will be addressed elsewhere. As shown in Fig. 1a, diagrammatically the
dressed Green function of atoms $G(\mathbf{k},i\omega _m)$ and its
self-energy $\Sigma _a(\mathbf{k},i\omega _m)$ read, respectively, 
\begin{eqnarray}
G^{-1}(\mathbf{k},i\omega _m) &=&G_0^{-1}\left( \mathbf{k},i\omega _m\right)
-\Sigma _a(\mathbf{k},i\omega _m),  \label{atomgf} \\
\Sigma _a(\mathbf{k},i\omega _m) &=&T\sum_{\mathbf{q},i\nu _n}\Gamma (\mathbf{q}%
,i\nu _n)G\left( \mathbf{q-k},i\nu _n-i\omega _m\right) ,  \label{atomse}
\end{eqnarray}
where $G_0\left( \mathbf{k},i\omega _m\right) =\left[ i\omega _m-\left(
\epsilon _{\mathbf{k}}-\mu \right) \right] ^{-1}$ is free atomic Green
function, and $i\omega _m$ ($i\nu _n$) is the fermionic (bosonic) Matsubara
frequencies. The approximate four-point vertex function $\Gamma (\mathbf{q}%
,i\nu _n)$ is solved readily from the Bethe-Salpeter equation (see Fig. 1c), 
\begin{equation}
\Gamma ^{-1}(\mathbf{q},i\nu _n)=U_{eff}^{-1}\left( \mathbf{q},i\nu
_n\right) +\Pi (\mathbf{q},i\nu _n),  \label{atomvf}
\end{equation}
where $\Pi =T\sum_{\mathbf{k,}i\omega _m}G\left( \mathbf{q-k},i\nu
_n-i\omega _m\right) G\left( \mathbf{k},i\omega _m\right) -\sum_{\mathbf{k}%
}1/\left( 2\epsilon _{\mathbf{k}}\right) $, and the effective pairing
interaction $U_{eff}=U_0+g_0^2D_0\left( \mathbf{q},i\nu _n\right) $ consists
of the background interaction $U_0$ and the induced interaction mediated by
Feshbach molecules $g_0^2D_0\left( \mathbf{q},i\nu _n\right) $, where $%
D_0\left( \mathbf{q},i\nu _n\right) =1/\left[ i\nu _n-\epsilon _{\mathbf{q}%
}/2-\nu _0+2\mu \right] $ is the non-interacting molecular Green function.
Note that the induced interaction dominates near the Feshbach resonance, and
acquires a dependence on the relative energy of the colliding atoms (\textit{%
i.e.}, retarded effects) that becomes stronger for a narrow resonance.
Analogously, it is straightforward to write down the atom-molecule vertex
function, 
\begin{equation}
g^{-1}\left( \mathbf{q},i\nu _n\right) =g_0^{-1}+g_0^{-1}U_0\Pi (\mathbf{q}%
,i\nu _n),  \label{molevf}
\end{equation}
whose diagrammatic equivalent is shown in Fig. 1d. This leads to the Dyson
equation for molecules (see Fig. 1b), 
\begin{eqnarray}
D^{-1}\left( \mathbf{q},i\nu _n\right) &=&D_0^{-1}\left( \mathbf{q},i\nu
_n\right) -\Sigma _m(\mathbf{q},i\nu _n),  \label{molegf} \\
\Sigma _m(\mathbf{q},i\nu _n) &=&-g_0\Pi (\mathbf{q},i\nu _n)g(\mathbf{q}%
,i\nu _n),  \label{molese}
\end{eqnarray}
where $D\left( \mathbf{q},i\nu _n\right) $ and $\Sigma _m(\mathbf{q},i\nu
_n) $ are, respectively, the molecular Green function and self-energy.

Eqs. (\ref{atomgf})-(\ref{molese}) form a close set of equations, which we
refer to as a self-consistent theory for the two-channel model. We emphasize
that in these equations, we have used a fully dressed fermionic Green
function that has to be determined self-consistently. This is contrasted to
the NSR scheme considered by Ohashi and Griffin, where a free fermionic
Green function is adopted \cite{griffin}. Actually, with replacing $G$ by $%
G_0$ in $\Pi $ and $\Sigma _a$ and writing $G=G_0+G_0\Sigma _aG_0$, we
reproduce the NSR formalism. Furthermore, if we set $U_{eff}\equiv U_0$ in
the four-point vertex function (\ref{atomvf}) and neglect the molecular
part, we recover the self-consistent theory for the single-channel model, as
proposed earlier by Haussmann \cite{haussmann}.

We have solved Eqs. (\ref{atomgf})-(\ref{molese}) in a self-consistent
manner by an iteration procedure, which is repeated successively to obtain $%
G $ and $D$ in higher orders until convergence is achieved. At a finite
temperature $T\geq T_c$, the only quantity to be determined is the chemical
potential $\mu $, which has to be adjusted to satisfy the constraint for the
total density $n_{tot}=n_F+2n_m$, where $n_F=2G(\mathbf{x=0},\tau =0^{-})$
and $n_m=-D(\mathbf{x=0},\tau =0^{-})$ are the density of atoms and
molecules, respectively. In the calculation, the energy and length are taken
in units of $\epsilon _F=\hbar ^2k_{F\text{ }}^2/2m$ and $k_{F\text{ }}^{-1}$%
, where $k_{F\text{ }}=\left( 3\pi ^2n_{tot}\right) ^{1/3}$ is the
characteristic Fermi wavelength. Without further clarity, in the calculation
we take a weak back-ground atom-atom interaction, $k_Fa_{bg}=-0.1$, in
accord with the real experimental situation \cite{hulet}.

It is worth noticing that the properties of Feshbach resonaces is naturally
characterized by a dimensionless atom-molecule coupling constant $g=g_0\hbar
^{-2}m\left( 2\pi k_{F\text{ }}\right) ^{-1/2}$, quoted later as an \emph{%
intrinsic} width of resonances. To illustrate the physical implication of
this dimensionless parameter, we note that the important energy scale of
resonances can be defined as $\epsilon _R=\Delta \mu \left( \Delta B\right)
_{in}$, where $\left( \Delta B\right) _{in}/\Delta B=\left( \Delta \mu
\Delta B\right) /\left[ \hbar ^2/\left( 2ma_{bg}^2\right) \right] $ \cite
{bruun,jasho}. $g$ is related to the ratio between $\epsilon _R$ and the
Fermi energy $\epsilon _F$ via $g=\left[ \epsilon _R/ \epsilon
_F \right] ^{1/4}$. Hence, if the Fermi energy is well within the
characteristic energy scale of resonances, \textit{i.e.}, $g\gg 1,$ we have
a broad resonance. The opposite limit of $g\leq 1$ defines a narrow one. We
note also that it is possible to tune the value of the intrinsic width of
resonances by changing the density of the gas, as $g\propto k_{F}^{-1/2}\propto n^{-1/6}_{tot}$.

\begin{figure}
\includegraphics[%
  width=7.5cm]{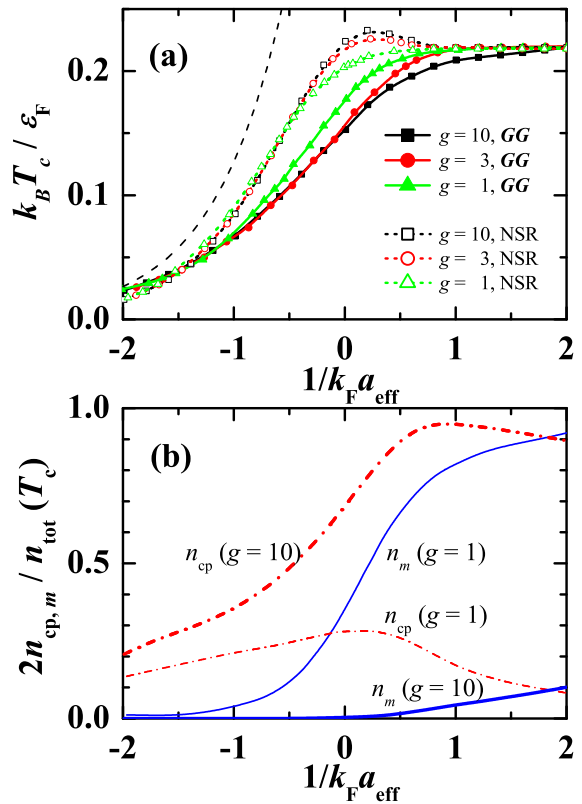}

\caption{(color online) (a) The superfluid transition temperature $T_c$ as a function of the
effective coupling strength $v_{eff}$ for dimensionless atom-molecule couplings $g=10,$ $%
3, $ and $1$. The full lines with solid symbols represent the numerical
results of our self-consistent theory. The dotted lines with empty symbols
are obtained by NSR method. (b) The population of bare Feshbach
molecules $n_m$ at $T_c$, and the population of preformed Cooper pairs $2n_{%
\text{cp}}=n_F-n_{F0}$ at $T_c$, where $n_{F0}=2\sum_{{\bf k}}f\left(
\epsilon _{{\bf k}}-\mu \right) $, and $f(x)$ is the Fermi distribution function.}

\label{fig2}
\end{figure}

\textit{Superfluid transition temperature $T_c$}. --- The superfluid
transition temperature can be conveniently determined by the Thouless
criterion,
\begin{equation}
\left[ \Gamma (\mathbf{q=0},i\nu _n=0)\right] _{T=T_c}^{-1}=0,
\label{thouless}
\end{equation}
which describes the instability of the normal phase of the gas towards the
formation of Cooper pairs of atoms \cite{nsr}. In our $T$-matrix
approximation, the Thouless criterion is consistent with the
Hugenholtz-Pines condition for Feshbach molecules that the molecular
Green-function (\ref{molegf}) must develop a pole at $\mathbf{q=}i\nu _n=0$
as $T\rightarrow T_c$. As the four-point vertex function represents roughly
the propagator of Cooper pairs, this consistency implies that the superfluid
phase transition is accomplished by dressed molecules that involve both the
preformed Cooper pairs and the Feshbach molecules \cite{griffin}.

In Fig. 2a we present our results for $T_c$ as a function of an effective
coupling constant $v_{eff}=1/\left( k_{F\text{ }}a_{eff}\right) $ as full
symbols, where $4\pi \hbar ^2a_{eff}/m=U_0+g_0^2/\left( -\nu _0+2\mu \right) 
$. For comparison the NSR predictions are also plotted by empty symbols.
Three values of dimensionless atom-molecule coupling $g$ have been
considered, corresponding to the situations with broad ($g=10$), medium ($g=3
$), and narrow ($g=1$) resonances. For all these intrinsic resonance widths,
it is obvious that our results of $T_c$ increase monotonically with
increasing the effective coupling constant. In contrast, for the broad or
medium resonance, the NSR theory predicts a slight hump structure at
the crossover. This maximum presumably is an artifact of the lack of
self-consistency. As stated by Haussmann \cite{haussmann}, a repulsive
interaction between the pairs leads to a depression of $T_c$ due to the
composite nature of pairs. The approximating use of the free fermionic Green
function in NSR scheme neglects the possible repulsion between dressed
molecules, and therefore overestimates the transition temperature.
Nevertheless, in two limits of weak and strong interactions, both theories
give almost the same results: In the weak-coupling limit of $%
v_{eff}\rightarrow -\infty $, they give back the standard BCS result of $%
k_BT_{BCS}/\epsilon _F\rightarrow 0.614\exp \left[ \pi v_{eff}/2\right] $,
as shown by the dashed line; On the other hand, in the strong-coupling limit
of $v_{eff}\rightarrow +\infty $, they correctly reproduce the BEC
temperature of $k_BT_{BEC}/\epsilon _F\approx 0.218$.

The transition temperature depends also on the intrinsic width of Feshbach
resonances in a non-trivial way. As shown in Fig. 2b, theoretically, with
decreasing the intrinsic width the character of dressed molecules changes
from the large object of Cooper pairs to the small point structure of
``bare'' Feshbach molecules. As a fact that the transition temperature of
non-interacting ``bare'' molecules is larger than that of performed Cooper
pairs, the less population of Cooper pairs tends to enhance our predicted
transition temperature. This feature should be universal with respect to the
inclusion of repulsive molecule-molecule interactions in the atom-molecule
model (\ref{hami}), because, contrasted to the composite Cooper pairs, the
transition temperature of a weakly interacting gas of \emph{true} molecules
increases with increasing repulsions between moleclues \cite{rmp}, and
thereby is always larger than that of Cooper pairs. We note that,
interestingly, the NSR results go exactly in the opposite direction as the
intrinsic width decreases, which is again due to the lack of
self-consistency.

\begin{figure}
\includegraphics[%
  width=7.5cm]{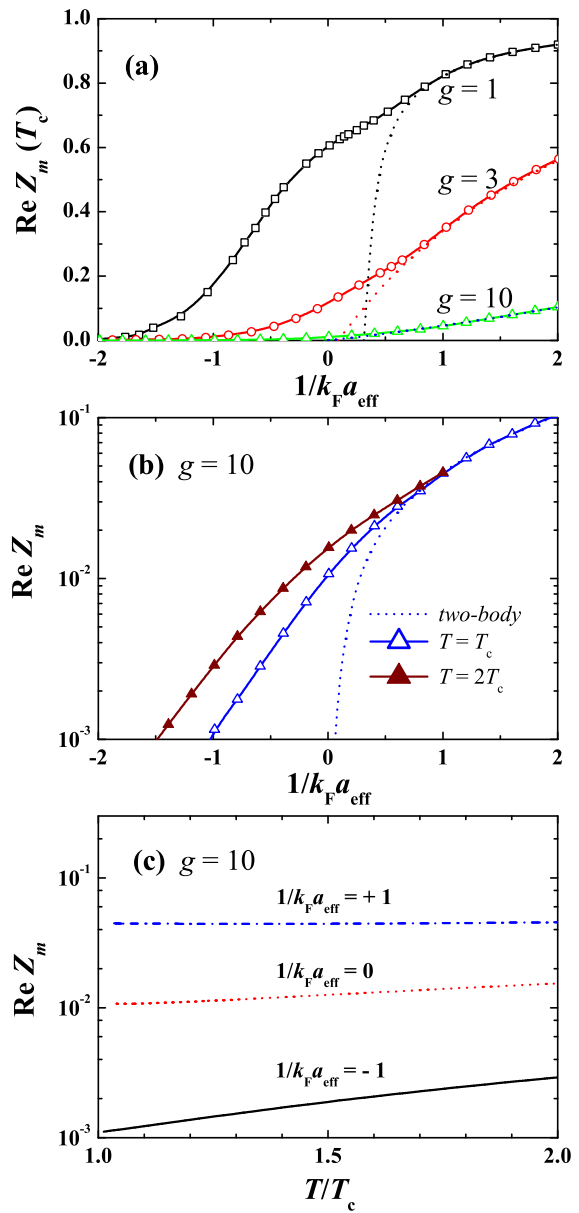}

\caption{(color online) (a) The real part of $Z_m$ at the superfluid transition as a function
of $v_{eff}$. (b) $\mathop{\rm Re}Z_m\left( T_c\right)$ and $\mathop{\rm Re}Z_m\left( 2T_c\right)$ 
at the crossover for $g=10$ in logarithmic scale. (c) Temperature dependence of 
$\mathop{\rm Re}Z_m$ at $v_{eff}=-1,$ $0,$ and $+1$ for $g=10$ .}

\label{fig3}
\end{figure}

\textit{Molecular probe of the BCS-BEC crossover}. --- At the crossover the
dressed molecules are extremely weakly bound objects, depending strongly on
many-body effects. They can hardly be detected by the standard technique
such as the time-of-flight expansion. In contrast, the ``bare'' Feshbach
molecules are well defined point structures, and can easily be characterized
by the usual optical spectroscopy methods. Therefore it is possible to probe
the crossover indirectly through the measurement of ``bare'' molecules. As
the dressed molecules contain both Cooper pairs and ``bare'' molecules \cite
{stoof}, \textit{i.e.}, $\left| \text{dressed}\right\rangle =e^{i\phi }\sqrt{%
1-Z_m}\left| \text{open}\right\rangle +\sqrt{Z_m}\left| \text{closed}%
\right\rangle $, a candidate observable would be the amplitude $Z_m$ of the
bare molecular state in the closed channel. For an ultracold gas of $^6$Li
atoms, the closed (open) channel, to an excellent approximation, corresponds
to the atomic singlet (triplet) state. As a result, the value of $Z_m$ can
directly be determined by selectively projecting out the singlet component
of the dressed molecules with optical spectroscopy. This is exactly what was
done recently by Hulet \textit{et al}. \cite{hulet}, where a tiny value of $%
Z_m$ ($\sim 10^{-4}$) has been observed in the unitary limit ($%
k_Fa_{eff}\rightarrow \infty $) right on resonance.

\begin{figure}
\includegraphics[%
  width=7.5cm]{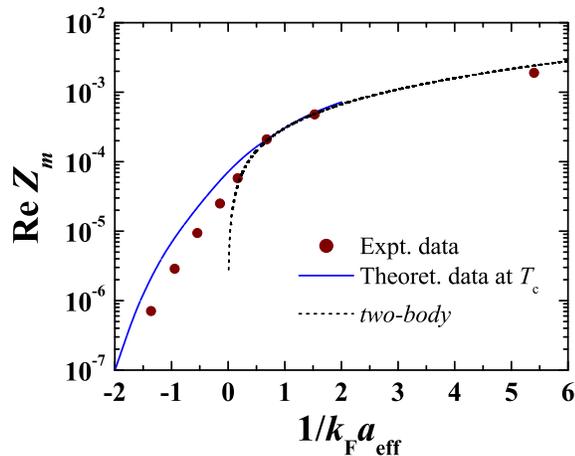}

\caption{(color online) Comparison between our theoretical predictions and the
experimental data (Fig. 5 in Ref. \cite{hulet}) for the residue of the pole of 
the $^6$Li Feshbach molecules at the broad resonance $B_0=834$ \textrm{G}.}

\label{fig4}
\end{figure}

In our theory $Z_m$ is associated with the residue of the pole of the
molecular Green function (\ref{molegf}), which, in the spirit of Landau
quasiparticle concept, takes the form, 
\begin{equation}
D\left( \mathbf{q},i\nu _n\right) =\frac{Z_m}{i\nu _n-\left( \hbar ^2\mathbf{%
q}^2/2m_m^{*}-\mu _m\right) },  \label{moleFR}
\end{equation}
for sufficiently small $\mathbf{q}$ and $i\nu _n$. Here $m_m^{*}$ is a
complex effective mass, and $\mu _m$ is a molecular chemical potential that
goes to zero as $T\rightarrow T_c$. In Fig. 3a, we report our
self-consistent calculations for $\mathop{\rm Re} Z_m$ at $T_c$ against the
coupling $v_{eff}$, together with the two-body results in dashed lines \cite
{drummond,stoof}. In the strong coupling BEC region ($v_{eff}\geq 1$) these
results are identical as expected. However, near the Feshbach resonance they
do show a large difference, which should attribute to the many-body effects.

The value of $Z_m$ decreases significantly with increasing the atom-molecule
coupling. In large $g$ limit where the Cooper pairs dominate at the
crossover, we find approximately $Z_m\left( g,v_{eff},T\right) \approx
Z_m\left( v_{eff},T\right) /g^2$. For $^6$Li gas at the broad resonance $%
B_0=834$ \textrm{G}, we estimate $g\approx 120$ \cite{hulet}. Therefore,
from our data at $g=10$, the expected value of $Z_m$ in the unitary limit is
in the order of $10^{-4}$, which agrees qualitatively well with the experimental
findings by Hulet \textit{et al}. \cite{hulet}. A full comparison between
our theorectial results and experimental data over the entire crossover regime is
shown in Fig. 4. We note that this qualitative agreement is robust against
the unknown temperature in experiment. In Figs. 3b and 3c, we show the finite
temperature effect on $Z_m$ in a logarithmic scale for $g=10$. As the
temperature doubles, $Z_m$ increases, but its order is essentially
unchanged. We note also that in our calculations we do not consider the
effect of harmonic trap present in experiments, which tends to
quantitatively decrease the value of $Z_m$.

\begin{figure}
\includegraphics[%
  width=7.5cm]{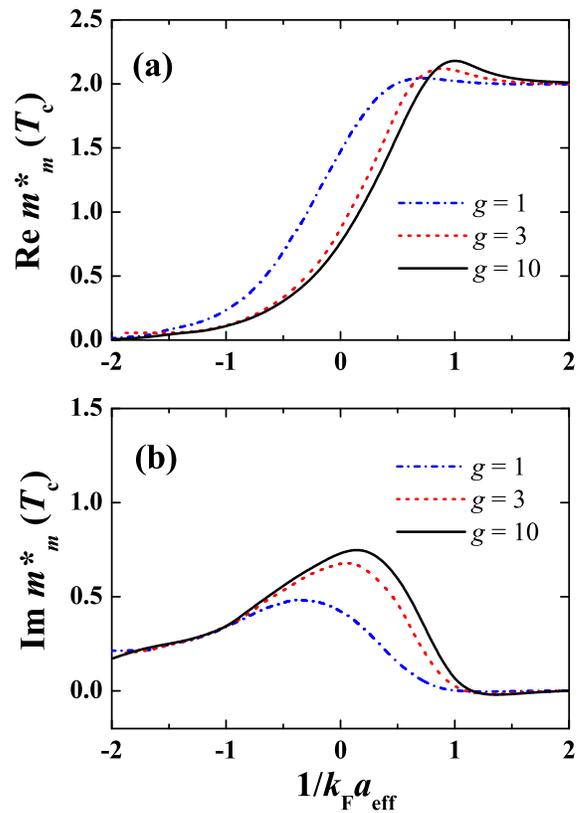}

\caption{(color online) The real part (a) and imaginary part (b) of the effective mass of the
bare molecules at the crossover.}

\label{fig5}
\end{figure}

At the crossover, the Feshbach molecules can transfer or break up into free
fermions. This mechanics causes the finite lifetime of the molecules, as
evidenced by the nonzero imaginary part of the residue $Z_m$, or of the
effective mass $m_m^{*}$. We present in Fig. 5 the results for $m_m^{*}$ at 
$T_c$. In the deep BEC limit, the molecules are stable so that the imaginary
part of $m_m^{*}$ is nearly zero and the real part $\mathop{\rm Re}m_m^{*}\approx
2m$. While in the opposite limit of a weak coupling, \textit{i.e.}, $v_{eff}\leq -1$, 
the imaginary part is much larger than the real part, which
suggests that the molecules are overdamped. In addition, as the
dimensionless atom-molecule coupling $g$ decreases, the value of 
$\mathop{\rm Im}m_m^{*}$ becomes smaller, as expected.

\begin{figure}
\includegraphics[%
  width=8.5cm]{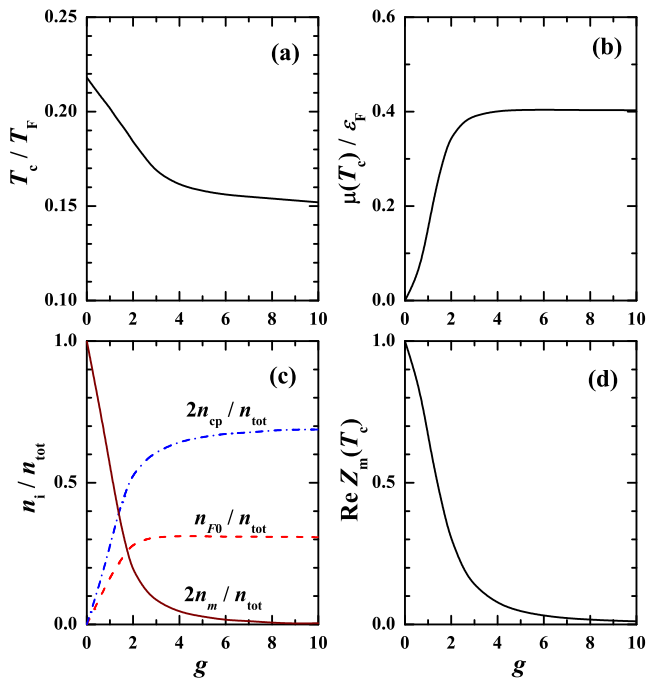}

\caption{(color online) Illustration of the crossover from narrow to broad resonances on
resonance $\nu _0=0$. (a) The transition temperature $T_c$, (b) the chemical
potential at $T_c$, (c) the fractions of particles, and 
(d) $\mathop{\rm Re}Z_m\left( T_c\right) $ as a function of the dimensionless 
atom-molecule coupling $g$.}

\label{fig6}
\end{figure}

\textit{Crossover from the narrow to broad resonances}. --- It was suggested
that for a broad Feshbach resonance, the gas should be well described by the
single-channel model with a single parameter (\textit{i.e.}, the $s$-wave
scattering length) \cite{bruun}. This suggestion was supported from a
coupled-channel calculation in the context of two-body physics \cite
{strinati}, or, from a mean-field calculation at zero temperature \cite
{jasho}. However, the situation is less clear beyond the mean-field
approximation. In Fig. 6, we show our many-body results on resonance ($\nu
_0=0$) for the critical temperature $T_c$, the chemical potential at $T_c$,
the residue of the pole of ``bare'' Feshbach molecules $Z_m$, as well as the
fractions of particles. As $g$ increases, $T_c$ and $\mu \left( T_c\right) $
come to saturate, and $Z_m$ and $n_m$ drop as $1/g^2$. Therefore, using our
self-consist theory we verified numerically that in the broad resonance
limit the behavior of the gas indeed becomes universal, towards the
prediction of the single-channel model.

A possible candidate to observe the crossover from narrow to broad
resonances is a gas of $^{40}$K atoms at $B_0=202$ \textrm{G}, where we find 
$g\approx 10$ under current experimental situation \cite{jin}. According to
the relation $g\propto n^{-1/6}$, it is possible to obtain a narrow
resonance by increasing the center density of the gas, $n$. This might be
achieved by using more tight harmonic traps or trapping more atoms in the
experiment.

\textit{Conclusions}. --- In this paper we have developed a many-body theory
of the two-channel model for an interacting Fermi gas with a Feshbach
resonance. The theory includes the strong fluctuations in a self-consistent
manner that is necessary at the BCS-BEC crossover. As an application, we
have studied in detail the normal phase properties of the crossover, with a
special attention on the effects of intrinsic resonance widths. A reasonable
superfluid transition temperature has been predicted. Our predictions for
the residue of the pole of ``bare'' molecules agree qualitatively well with
the experimental findings. We expect, therefore, that the present theory
gives a reliable description of current BCS-BEC crossover physics in atomic
normal Fermi gases. The generalization of our theory to the superfluid phase
is under investigation, and will be reported elsewhere.

We are indebted to Prof. Peter D. Drummond and Prof. Allan Griffin for useful
discussions. We gratefully acknowledge the Australian Research Council for
the support of this work.

\end{document}